\DeclareRobustCommand\openone{\leavevmode\hbox{\small1\normalsize\kern-.33em1}}
\newcommand{\oper}[2]{\hat{#1}^{\phantom{\dag}}_{\bf #2}}

\newcommand{\operc}[1]{\hat{c}^{\phantom{\dag}}_{\textbf #1}}
\newcommand{\operdagc}[1]{\hat{c}^{\dag}_{\textbf #1}}
\newcommand{\annihilator}{\operc{i \sigma}}
\newcommand{\creator}{\operdagc{i \sigma}}
\newcommand{\numberop}[1]{\oper{n}{#1}}
\newcommand{\numberup}{\numberop{i \uparrow}}
\newcommand{\numberdown}{\numberop{i \downarrow}}
\newcommand{\numberi}{\numberop{i}}

\newcommand{\veci}{\textbf{i}}

\newcommand{\vecS}{\textbf{S}}

\documentclass[aps,prl,twocolumn,groupedaddress,showpacs]{revtex4}

\usepackage{graphicx,amsmath,amssymb,color}
\usepackage{hyperref}

\begin{document}

\title{DMFT vs Second Order Perturbation Theory for the Trapped 2D Hubbard-Antiferromagnet}

\author{Andreas D. Pfister}
\author{Eberhard Jakobi}
\author{Tobias Gottwald}
\email{tobias.gottwald@uni-mainz.de}
\author{Peter\ G.\ J. \surname{van Dongen}}

\affiliation{KOMET 337, Institut f\"ur Physik, Johannes Gutenberg-Universit\"at, Mainz}

\date{\today}

\begin{abstract}
In recent literature on trapped ultracold atomic gases, calculations for 2D-systems are often done within the Dynamical Mean Field Theory (DMFT) approximation. In this paper, we compare DMFT to a fully two-dimensional, self-consistent second order perturbation theory for weak interactions in a repulsive Fermi-Hubbard model. We investigate the role of quantum and of spatial fluctuations when the system is in the antiferromagnetic phase, and find that, while quantum fluctuations decrease the order parameter and critical temperatures drastically, spatial fluctuations only play a noticeable role when the system undergoes a phase transition, or at phase boundaries in the trap. We conclude from this that DMFT is a good approximation for the antiferromagnetic Fermi-Hubbard model for experimentally relevant system sizes.
\end{abstract}


\maketitle

\section{Introduction}
The study of ultracold, trapped atomic gases on a lattice as an emulation of tight-binding Hamiltonians has been an active field of research for the past 15 years.
After the realization of BECs of alkali atoms in 1995 \cite{Davis1995,Anderson1995}, many theoretically predicted quantum phenomena, such as the bosonic superfluid-Mott insulator transition\cite{Greiner2002} and superexchange\cite{Trotzky2008}, have been observed experimentally. The recent achievement of a fermionic Mott insulator in a Hubbard Hamiltonian\cite{Jordens2008,Schneider2008} has steered the focus towards understanding the fermionic Hubbard model and its phase diagram, which may have a relevance for high temperature superconductivity\cite{Anderson1987,Andersen2007}. An important challenge is to reach the N\'eel temperature, in order to realize antiferromagnetic phases. \par

It is common practice, when doing calculations on two-dimensional trapped atomic gases, to use DMFT\cite{Snoek2011}, which includes only spatially local fluctuations. While DMFT is known to be a good approximation in three dimensional problems\cite{Georges1996,Kotliar2004}, it is expected to perform much more poorly in two dimensions, since non-local fluctuations gain importance\cite{Schweitzer1990,Schweitzer1991}. \par

We have therefore implemented a self-consistent second order perturbation theory for the Fermi-Hubbard Hamiltonian. The self energy expansion, which is described further on, includes local as well as non-local fluctuations of the system. This gives us the tools to compare the effects of quantum and spatial fluctuations for the model, up to second order in the interaction. \par

A two pseudospin-species Fermi mixture in a magneto-optical trap realizes a Hubbard model with broken lattice symmetry,
\begin{multline}
 \mathcal{H} = -t \sum_{(\bf{ij} ), \sigma} \operdagc{i \sigma} \operc{j \sigma} + U \sum_{\veci} \numberup \numberdown \\
 + \sum_{\veci \sigma} \left( V{\veci}^{2} - \mu_{\sigma} \right) \numberop{i \sigma} \; ,
\end{multline}
where $( \bf ij )$ denotes the sum over next neighbors, $ \creator $ and $\annihilator $ are the creation and annihilation operators for the respective lattice point and spin and $\numberop{i \sigma}=\creator\annihilator$ is the number operator.
Also, $t$ is the next-neighbor hopping amplitude, and $U$ the on-site interaction, which is chosen to be repulsive, $U>0$, in this paper. 
The term $V{\bf i}^{2}$ describes the harmonical trapping potential which breaks the lattice translational symmetry, and the spin-dependent chemical potential is $\mu_{\sigma}$.
The energy scale for this work is the hopping amplitude, $t$ ($t\equiv 1$).\par

It is necessary to use approximative schemes to solve this Hamiltonian. In this paper, three different approximations are compared, namely Hartree-Fock theory, DMFT, and self-consistent perturbation theory up to second order in the interactions.
It is known from Hartree-Fock theory\cite{Gottwald2009}that, for the case of an imbalance, $\mu_{\uparrow}\neq\mu_{\downarrow}$, the ground state is a canted antiferromagnet, with a $U(1)$-symmetry for all rotations around the z-axis.
For the case of balanced spin species, $\mu_{\uparrow}=\mu_{\downarrow}$, the system is symmetric under all rotations of SU(2). While the code used can easily reproduce both balanced and imbalanced cases, this paper will be constrained to balnced systems, for simplicity. These are sufficient for an identification of key effects of second order diagrams.\par

This paper is organized as follows. First, we will give a short introduction to the approximation schemes used here. Then, we will discuss results for the trapped system, focusing on the differences of the approaches. The role of non-local diagrams is then investigated further, and finally a short survey of finite size effects, which are noticeable in our calculations, is presented.\par

\section{The Different Approximative Schemes}
\begin{figure}[htb]
 \centering
 \includegraphics[height=0.35\textwidth,angle=-90]{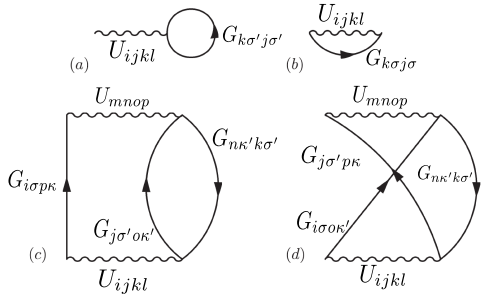}
 \caption{First and second order SE diagrams. a) Hartree-term b) Fock-term c) density-density term d) exchange term}
 \label{fig:diags}
\end{figure}

In \textbf{Hartree-Fock theory} (HF), quantum fluctuations are completely neglected. This is expressed through the transformation
\begin{equation}\label{Hartree-Fock}
 \numberup \numberdown \rightarrow  2 \langle \numberi \rangle \numberi - 2 \langle \oper{\vecS}{i} \rangle \cdot \oper{\vecS}{i} - \langle \numberi \rangle^2 + \langle \oper{\vecS}{i} \rangle^2 \; .
\end{equation}
This is equivalent to a self-consistent perturbation scheme of first order [c.f. Figs. \ref{fig:diags} a) and b)], with a local and frequency-independent self energy (SE), which is therefore diagonal in real space representations:
\begin{equation}
 \Sigma_{i\sigma,j\kappa}(\omega) = \Sigma^{HF}_{i\sigma,j\kappa}\delta_{ij} \; .
\end{equation}
The neglect of quantum fluctuations leads to a systematical exaggeration of both critical temperature and ground state staggered magnetization in all dimensions.\par

\textbf{Dynamical Mean Field Theory} (DMFT) tries to correct for this shortcoming by including some quantum fluctuations, namely all local ones. The resulting self energy continues to be diagonal in real space, but now has a frequency dependence:
\begin{equation}\label{dmft}
 \Sigma_{i\sigma,j\kappa}(\omega) = \left[\Sigma^{HF}_{i\sigma,j\kappa} + \Sigma^{DMFT}_{i\sigma,j\kappa}(\omega)\right]\delta_{ij} \; .
\end{equation}
DMFT is usually viewed as the limit of infinite dimensions, and as such is a useful method for high ($d\ge 3$) dimensions.
It is generally expected, however, that in lower dimensions the usefulness of DMFT is strongly reduced, because nonlocal processes gain in weight.
DMFT is a priori a nonperturbative expansion: the SE includes diagrams of \emph{all} orders of perturbation. 
For the sake of our comparison, we have only included contributions to the local SE up to \emph{second order}. We expect this to deviate only slightly from the full DMFT, for the same reasons given below for full second order calculations.
In the diagrammatic language, this corresponds to an inclusion of diagrams a) to d) in Fig. \ref{fig:diags}, but only those whose Green functions' lattice indices are equal, $G_{i\sigma, i\kappa}$ for all Green functions in the diagram.\par

Finally, the \textbf{self-consistent second order perturbation theory} (2OPT) includes local and nonlocal diagrams. The SE is now nondiagonal in real space,
\begin{equation}\label{o2PT}
 \Sigma_{ij}(\omega) = \Sigma^{HF}_{i\sigma,j\kappa}\delta_{ij} + \Sigma^{2OPT}_{i\sigma,j\kappa}(\omega) \; .
\end{equation}
The numerical effort grows exponentially with every order of interactions in the perturbative series, making it numerically difficult to include more than second order diagrams. 
Because we stay in the weak interaction regime, and higher orders of interactions are suppressed by at least $U^3/W^2$ -- where $W=4t$ is the typical half-bandwidth of the two-dimensional Hubbard model -- second order calculations should include all dominant quantum fluctuations and, therefore, return quantitatively correct results.

\section{The Inhomogeneous System}

\begin{figure}[htbp]
 \centering
 \includegraphics[width=0.482\textwidth]{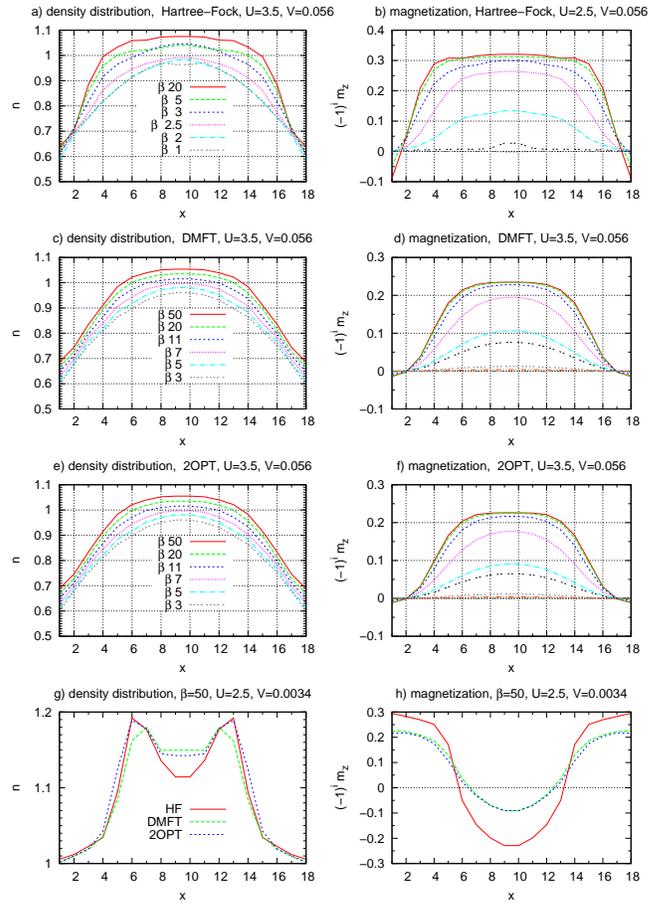}
 \caption{(Color online) Cuts through the lattice for different approximations, displaying the occupation and the staggered z-magnetization. Figs. a) to f) show almost the entire occupied region, with trap potential $V=0.56$, while  Figs. g) and h) show the center of a system, with $V=0.0017$. The HF results show significantly higher critical temperatures and higher ground state staggered magnetizations, as well as sharper phase transitions, than the other two approximations, 2OPT and DMFT, which differ only slightly, although consistently, from another.}
 \label{fig:trap_dens}
\end{figure} 

For the Fermi-Hubbard model on a square lattice, the additional trapping potential leads to a coexistence of different phases in the system, depending on the local density: For local densities lower and significantly higher than half filling, a paramagnetic phase is present. In regions close to half filling, a N\'eel state is present. While at high temperatures the N\'eel state adapts to the filling as given by the trap potential, when temperatures are lowered the N\'eel phase enforces half filling over a broad range, reminiscent of the occurrence of phase separation in a doped homogeneous system\cite{Dongen1995}. Unlike the Mott-plateau, which is a fluctuation effect not reproducible in HF-calculations \cite{Helmes2008}, the N\'eel-plateau can be seen even at HF-level.

A domain wall boundary appears as soon as  the system crosses into the N\'eel-plateau region \cite{Gottwald2009}. At the new boundary, the antiferromagnetic order parameter $(-)^\veci m_z(\veci)$ changes its sign. The boundary can also be seen in the occupation number as a ring of higher occupation extending from the region of half filling [Figs. \ref{fig:trap_dens} g), h)]. Notice that the higher occupation is suppressed again towards half filling by the second antiferromagnetic domain, working in opposition to the trapping potential. This new N\'eel domain requires the occupied part of the lattice to extend to a radius of at least 15 to 20 lattice points around the trap center, otherwise the lattice resolution is too small for the new domain to exist.\par

Our calculations show that quantum fluctuations increase the inverse N\'eel temperature, $\beta_N$, while decreasing the N\'eel order parameter, c.f. Fig. \ref{fig:trap_dens}. In HF, $\beta_N^{HF}(U=3.5)$ is $1.5$, while $\beta_N^{DMFT}(U=3.5)$ and $\beta_N^{2OPT}(U=3.5)$ are $3$. The (effectively) ground-state staggered magnetization at $U=7$ is reduced from $\Delta^{HF}(\beta=30)=0.319$ to $\Delta^{DMFT}(30)=0.235$ and $\Delta^{2OPT}(30)=0.226$, by 30\%. The progression from low to high temperatures can be seen in Figs. \ref{fig:trap_dens} a) to f). In HF, the magnetization remains sizeable up to comparatively high temperatures, and the crossover to the low-T N\'eel state occurs at $\beta_c^{HF}=4$ [c.f. Fig. \ref{fig:trap_dens} a), where the plateau disappears for high-T calculations]. In  Figs. \ref{fig:trap_dens} c) to f), the results for DMFT and 2OPT are shown. The crossover behavior is smeared out over a wide range of temperatures, starting to set in at $\beta=8$ and being fully developed only around $\beta=18$.\par

\section{The Role of Non-local Diagrams}

\begin{figure}[htbp]
 \centering
 \includegraphics[width=0.4\textwidth]{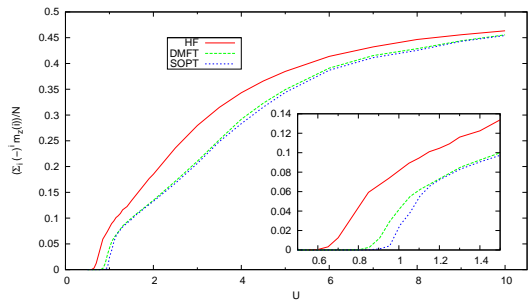}
 \caption{(olor online) Comparison of HF to 2OPT and DMFT calculations on the $U$-axis. The system is a 16 by 16 lattice, with periodic boundaries at half filling and low temperatures ($\beta=50$). 2OPT and DMFT differ only marginally.}
 \label{fig:o1o2dmft}
\end{figure}

In this section, we want to concentrate further on the conformance of 2OPT and DMFT results. To this end, we take a look at finite systems with periodic boundary conditions, leaving out the trapping potential, $V=0$, for translational invariance. This gives access to the effects of local and non-local diagrams without parallel effects from trap curvature or boundary potentials. \par

Fig. \ref{fig:o1o2dmft} offers a typical progression of 2OPT corrections to the systems: At very weak interactions, no order is visible. The critical interaction $U_N^{HF}(\beta)$ is reached in HF before $U_N^{DMFT}(\beta)$ is reached in DMFT, which in turn is closely followed by $U_N^{2OPT}(\beta)$. In medium-low interaction ranges, the suppression reaches a maximum, usually around $2<U<3$, and then slowly weakens at higher interactions. This last piece of information must, however, be treated with caution, as our method is of limited validity at higher interactions.\par

\begin{figure}[htbp]
 \centering
  \includegraphics[height=0.4\textwidth,angle=-90]{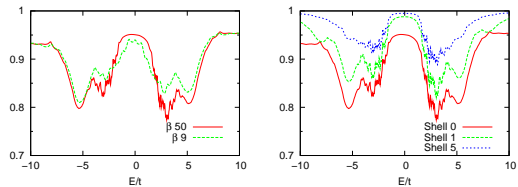}
 \caption{(Color online) Ratio of local to non-local elements in the self energy field. Fig. a) shows the ratio for two different temperatures $T_1=0.02$, which is strongly antiferromagnetic, and $T_2=0.11$, which is only very weakly magnetized. The weight of non-local elements never exceeds about 20\%, and sinks to about 5\% in an area around the Fermi energy. Fig. b) shows a selection of the weight of different shells in real space around purely local elements: Shell 1 thus means local and next-neighbor self energy elements, shell 5 all elements with exchange between lattice site up to 5 jumps (Manhattan metric) from the original site.}
 \label{fig:se_ratio}
\end{figure}

Again in Fig. \ref{fig:o1o2dmft}, the close proximity of 2OPT and DMFT results when sampling systems with different interaction strength $U$ at constant temperature and filling can be seen. While DMFT is slightly above 2OPT for the entire graph, the difference is only 1-4\% at interactions from $U=1.25$ onwards. The only clear deviation is near the critical interaction $U_N$, where 2OPT shifts the critical interaction further from $U_N^{HF}(\beta)$ than DMFT.\par

The close agreement of the data can also be seen in the self energy. To ascertain which role is played by non-local fluctuations, we compared the weight of local and non-local self energy diagrams in a 2OPT calculation. The calculation was done by adding the moduli of all local self energy matrix elements, and normalizing to the summed moduli of all self energy diagrams for each frequency. The result, shown in Fig. \ref{fig:se_ratio}, is that off-diagonal elements contribute only weakly to the antiferromagnet. Especially around the Fermi frequency, the local diagrams make up 95\% of the self energy's weight. The picture is essentially the same for high and low temperatures, and for different relevant interaction strengths.\par

We conclude from these graphs that, for antiferromagnetic systems, the influence of non-local fluctuations is weaker than generally expected, and that DMFT is a valid approximation in the trapped weak coupling 2D-antiferromagnet. The only notable divergences are near phase boundaries, both spatial boundaries, as in Figs. \ref{fig:trap_dens} g) and h), or in parameter space, as in Fig. \ref{fig:o1o2dmft}. \par

We were able to investigate lattice sizes of up to 18 by 18 atoms. Since this is smaller than common experimental extents, we include an investigation on finite size effects on the system in this paper. For these runs, we again chose periodic boundary conditions and no harmonic potential, in order to investigate only effects coming from the increasing number of nonlocal diagrams and lattice points, and eliminate effects stemming from different curvature of the trapping potential in the discretized lattice, and from the constant boundary condition of the experimental setup.\par

\begin{figure}[htb]
 \includegraphics[height=0.482\textwidth,angle=-90]{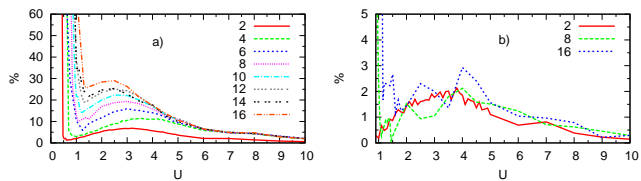}
 \caption{(Color online) a) The relative suppression of the antiferromagnetic phase through second order diagrams as compared to HF results. The labels refer to the system size, and all calculations are done at $\beta=50$ and half filling in a periodic system. The larger the system, the stronger the suppression.\ b) The deviation of DMFT and 2OPT results for exemplary system sizes. Shown are the regions where both approximations have appreciable staggered magnetization. For all system sizes investigated, no noticeable trend can be made out.}
 \label{fig:sup}
\end{figure}

The effect of finite system sizes is pronounced, as can be seen in Fig. \ref{fig:sup} a): The critical interaction $U_N(\beta)$ is shifted by 2OPT; the larger the system the stronger the shift. The amplitude of the suppression also increases with the system size, so that larger systems reduce antiferromagnetic order, as expected from the Mermin-Wagner theorem. As a direct consequence of this, we expect experimental results to show a stronger reduction of antiferromagnetic order, and thus a lower critical temperature, than what we have calculated here.\par

In Fig. \ref{fig:sup} b), we show the relative deviation of DMFT to 2OPT runs for different system sizes. No trend in lattice extents can be made out. We deduce, that the spatial fluctuations converge quickly for increasing shells around the local approximation, and do not change appreciably for system sizes of more than 10 by 10. All finite size effects in Fig. \ref{fig:sup} a) thus stem from local fluctuations.


\section{Conclusions}
In summary, we have investigated the antiferromagnetic phase of ultracold atoms trapped on a lattice in second order self-consistent perturbation theory, and compared these results to HF and DMFT approaches.

We find that the inclusion of quantum fluctuations to HF calculations strongly shifts critical temperatures and order parameters to lower values, as expected. Inclusion of non-local parts to the self energy changes little over a large range of values. Only near critical points and boundaries do noticeable variations appear between second order perturbation theory and dynamical mean field theory. This is reflected in the small weight of non-local elements in the self energy. \par

We then looked at the progression of the suppression with increasing lattice size. The suppression through quantum fluctuations increases monotonically with the size, showing pronounced finite size effects. Spatial fluctuations, on the other hand, quickly converge at small system sizes, and are fully represented in the systems we investigate.

\bibliography{ref}

\end{document}